\documentclass[aip,jcp,reprint,floatfix,twocolumn,citeautoscript]{revtex4-2}

\usepackage{amssymb}
\usepackage{amsmath}
\usepackage{amsfonts}
\usepackage{graphicx}
\usepackage{hyperref}

\renewcommand{\vec}[1]{\boldsymbol{#1}}

\newcommand{\half}{\tfrac{1}{2}}

\newcommand{\rcite}[1]{Ref.~\onlinecite{#1}}

\newcommand{\rcites}[1]{Refs~\onlinecite{#1}}

\newcommand{\Hrm}{\text{H}}
\newcommand{\xc}{\text{xc}}

\newcommand{\DFA}{\text{DFA}}

\newcommand{\pr}{^{\prime}}

\renewcommand{\vr}{\vec{r}}

\newcommand{\vrp}{\vec{r}\pr}

\newcommand{\up}{\mathord{\uparrow}}
\newcommand{\down}{\mathord{\downarrow}}

\newcommand{\Err}{\text{Err}}
\newcommand{\ext}{\text{ext}}
\newcommand{\MF}{\text{MF}}
\newcommand{\sys}{\text{sys}}

\usepackage[normalem]{ulem}
\usepackage{color}
\usepackage{xcolor}

\newcommand{\comment}[1]{}

\begin{document}

\title{A step toward density benchmarking --
the energy-relevant ``mean field error''}

\author{Tim Gould}\affiliation{Qld Micro- and Nanotechnology Centre, %
Griffith University, Nathan, Qld 4111, Australia}

\begin{abstract}
Since the development of generalized gradient
approximations in the 1990s, approximations based
on density functional theory have dominated
electronic structure theory calculations.
Modern approximations can yield energy differences
that are precise enough to be predictive in many instances,
as validated by large- and small-scale benchmarking
efforts.
However, assessing the quality of densities has been
the subject of far less attention, in part because
reliable error measures are difficult to define.
To this end, this work introduces the mean-field error
that directly assesses the quality
of densities from approximations.
The mean-field error is contextualised within
existing frameworks of density functional error
analysis and understanding, and
shown to be part of the density-driven error.
It is demonstrated on several illustrative examples.
Its potential use in future benchmarking protocols
is discussed, and some conclusions drawn.
\end{abstract}

\maketitle

\section{Introduction}
\label{sec:Intro}

Modern density functional theory (DFT) was introduced
in the mid 1960s through the pioneering work of
Hohenberg, Kohn and Sham.~\cite{HohenbergKohn,KohnSham}
Since the 1990s it has come to dominate computational
electronic structure theory, with tens of thousands
of works published each year.
Calculations based on DFT are routinely used to
understand chemical and solid state problems.
Increasingly, they are used to predict the properties
of novel molecules and materials.

The reason for DFTs dominance can mostly be
explained by two important elements of the approach:
1) approximations based on the one-body density
(density functional approximations)
can perform remarkably well as a tool for prediction
and analysis of quantum chemical problems;
2) the use of one-body densities as the key variable
of interest dramatically reduces computational
demands compared to many-body wave functions, and makes
the study of large systems numerically tractable.
With these elements in mind, it is time to
move on to the theoretical basis of DFT and its
approximations.

Consider a molecular or solid state system defined
by a nuclear potential, $v_{\sys}$. The Hohenberg-Kohn
and Kohn-Sham theories~\cite{HohenbergKohn,KohnSham}
dictate that its ground state energy may be
written as a functional,
\begin{align}
E_{\sys}=E[n_{\sys}] =&F_{\MF}[n_{\sys}]
+ E_{\xc}[n_{\sys}]\;,
\label{eqn:Esys}
\end{align}
of its optimal electron (one-body) density, $n_{\sys}$.
Eq.~\eqref{eqn:Esys} somewhat unconventionally
divides the usual DFT energy into a mean-field part,
\begin{align}
F_{\MF}[n]=&T_s[n] + \int n v_{\sys} d\vr
+ E_{\Hrm}[n]
\label{eqn:FMF}
\end{align}
and the usual exchange-correlation (xc)
energy, $E_{\xc}[n]$.
The mean-field part is a semi-classical approximation
to the physics of the system, in which the only
quantum effects are in the KS kinetic energy,
$T_s[n]$, whereas interactions of electrons with
nucleii, $\int n v_{\sys} d\vr$, and each other,
$E_{\Hrm}[n]=\half \int n(\vr) n(\vrp)\tfrac{d\vr d\vrp}{|\vr-\vrp|}$,
are treated classically.
All other quantum contributions are in the
unknown exchange correlation (xc) energy,
$E_{\xc}[n]$.

All the ingredients of $F_{\MF}$ are known in
an orbital formalism, so may be computed exactly.
The remarkable practical success of DFT comes from
the fact that $E_{\xc}[n]$ may be replaced by a
density functional approximation (DFA),
$E_{\xc}^{\DFA}[n]$, yet still yield useful
energetic predictions.
The ground state energy from the DFA is,
\begin{align}
E_{\sys}^{\DFA}=E^{\DFA}[n_{\sys}^{\DFA}]
=&F_{\MF}[n_{\sys}^{\DFA}]
+ E_{\xc}^{\DFA}[n_{\sys}^{\DFA}]\;,
\label{eqn:EsysDFA}
\end{align}
where all terms are easily computable.
The density, $n_{\sys}^{\DFA}$, is found by
self-consistently solving,
\begin{align}
\big\{
-\half\nabla^2 + v_{\ext} + v_{\Hrm}[n] + v_{\xc}^{\DFA}[n]
\big\} \phi_k = \epsilon_k \phi_k
\end{align}
to obtain $n_{\sys}^{\DFA}=\sum_k f_k|\phi_k|^2$
and $T_s[n_{\sys}^{\DFA}]=\half \sum_k f_k \int |\nabla\phi_k|^2 d\vr$.
Here,
$v_{\Hrm}=\int n(\vrp)\tfrac{d\vrp}{|\vr-\vrp|}$ is
the Hartree potential  and
$v_{\xc}^{\DFA}=\delta E_{\xc}^{\DFA}/\delta n$
is the xc potential.

Given the importance of energies in physical systems,
typically the quality of any DFA is assessed on how well
(or not) it reproduces energy properties (usually energy
differences). The magnitude of energy errors,
\begin{align}
\Err_E=&|E_{\sys}^{\DFA} - E_{\sys}|
=|\text{Eq.~\eqref{eqn:EsysDFA}}-\text{Eq.~\eqref{eqn:Esys}}|\;,
\end{align}
has therefore
been the subject of significant scrutiny through
large (e.g. \rcites{Karton2011,Mardirossian2017,Goerigk2017,Chan2019}
and references therein)
and boutique (e.g. \rcites{Tawfik2018,Seeger2020,Mester2022})
benchmarking efforts.
Thirty years of work on generalized gradient
approximations (GGAs) and higher rungs of
Jacob's ladder~\cite{Perdew2001-Jacob}
have led to DFAs that can (semi-)\cite{Gould2022-Poison}
reliably predict reaction energies to within
10~kcal/mol.~\cite{Mardirossian2017,Goerigk2017}
In many cases, better predictions have been enabled
by heavily empirical optimization strategies that
eschew physical constraints for lower errors on
training sets.

However, comparing Eqs~\eqref{eqn:EsysDFA} and
\eqref{eqn:Esys} reveals that the DFA involves
\emph{two} approximations: one for the energy,
$E_{\xc}\to E_{\xc}^{\DFA}$, and one for the
density, $n_{\sys}\to n_{\sys}^{\DFA}$.
One might expect that as DFAs improve in quality,
and as they climb Jacob's ladder,~\cite{Perdew2001-Jacob}
that their energies and densities should both get
better as they navigate the path toward exactness.
However, in ground breaking work, Medvedev et al~\cite{Medvedev2017}
reported that the empricial strategy seems to have led to
better energies at the expense of worse densities
-- leading DFAs to ``stray from the path'' to exactness.

This apparent decoupling of the quality of energies
and densities has consequently led to increasing
interest (e.g. \rcites{Kepp2017,Gould2017-Fukui,Mezei2017,Hait2018,Bremond2021,LanderosRivera2022,Sim2022,Dasgupta2022}) 
in the quality of DFA densities. That is, on the
magnitude of errors,
$\text{Err}[n_{\sys}^{\DFA}-n_{\sys}]$,
according to some quality measure.

Despite increasing interest, the scope of efforts to
benchmark densities is dwarfed by traditional
analyses based on energies (the author has seen
fewer papers on density benchmarking than there
are unique benchmarking \emph{sub}sets in
\rcite{Mardirossian2017} or \onlinecite{Goerigk2017}).
A major outstanding problem hampering better
understand of density errors is the lack of reliable
(i.e. useful and meaningful expressions) measures
for $\text{Err}[n_{\sys}^{\DFA}-n_{\sys}]$.
This work seeks to address this
problem by motivating and introducing the
``mean-field error'' that assesses densities
according to their DFA-independent effect on
energies.

The rest of the manuscript proceeds as follows.
First, the challenge of assessing densities is
discussed, and the mean-field error motivated.
Then, methodology is introduced and used
to investigate mean-field (and other)
measures of density errors on same exemplar
systems.
Next, some the future of density benchmarking
is discussed, wherein results from this work are
used to highlight some promissing leads and
highlight challenges that most be overcome
for routine benchmarking of densities.
Finally, some conclusions are drawn.

\section{Assessing errors in DFAs}
\label{sec:Errors}

Assessing energies of DFAs is straightforward,
albeit not uncontentious.~\cite{Gould2022-Poison,Weymuth2022}
One typically decides on a relevant set (typically between
20 and a couple of thousand examples )
of energy differences
-- such as a dissociation curve, a reaction energy,
the interaction between dimers or
the atomisation energy of a solid
-- and then evaluates how accurately one or more DFAs
reproduce them on average. By choosing energy differences
one is able to separate \emph{systematic} errors
in energies (which cancel out in differences)
from non-systematic errors, which are relevant for
practical computational studies. 

The simplest example of an energy difference metric is,
\begin{align}
\Err_E =&|\Delta E_{\sys}-\Delta E_{\sys}^{\DFA}|\;.
\label{eqn:ErrE}
\end{align}
Here,
\begin{align}
\Delta E_{\sys}=& E_{\sys} - \sum_{A\in\sys}E_{A}\;,
\end{align}
is an energy difference in the true system, and,
\begin{align}
\Delta E_{\sys}^{\DFA}=& E_{\sys}^{\DFA} - \sum_{A\in\sys}E_{A}^{\DFA}\;,
\end{align}
is its DFA counterpart, where
$A$ indicates a subsystem of $\sys$,
whether that be constituent atoms are other
chemically relevant divisions.
A more general expression
is $|\sum_R w_{\sys_R} (E_{\sys_R} - E_{\sys_R}^{\DFA})|$ where
$w_{\sys_R}$ is the weight (generally a positive or negative
integer) assigned to some reactant, $\sys_R$.

By contrast, densities are non-local quantities
and there are a multitude of potential metrics
and measures (here meaning useful expressions that
do not meet the requirements of a metric)
for assessing them. Some
examples include:
\begin{itemize}
\item The mean absolute density metric,
\begin{align}
\Err_{\text{D}} =& \int |n_{\sys}^{\DFA}(\vr) - n_{\sys}(\vr)|d\vr\;,
\label{eqn:ErrD}
\end{align}
that has units of inverse volume;
\item The root mean square density metric,
\begin{align}
\Err_{\text{D}'} =& \sqrt{\int |n_{\sys}^{\DFA}(\vr) - n_{\sys}(\vr)|^2d\vr}\;,
\end{align}
that also has units of inverse volume;
\item The dipole difference measure,
\begin{align}
\Err_{\text{Dip.}} =& \bigg|\int
\big[n_{\sys}^{\DFA}(\vr) - n_{\sys}(\vr) \big] \vr d\vr\bigg|\;,
\end{align}
that has units of length;
\item The Hartree metric,
\begin{align}
\Err_{\Hrm} =& E_{\Hrm}[ n_{\sys}^{\DFA} - n_{\sys} ]
\label{eqn:ErrH}
\end{align}
that has units of energy.
\end{itemize}

These definitions may also be extended to
differences, e.g., the Hartree error
equivalent of eq.~\eqref{eqn:ErrE} is,
\begin{align}
E_{\Hrm}\big[ n_{\sys}^{\DFA} - n_{\sys} - {\textstyle\sum_A} (n_A^{\DFA}-n_A) \big]\;,
\end{align}
which similarly avoids systematic errors; while,
$E_{\Hrm}[ \sum_{R}w_{\sys_R} (n_{\sys_R}^{\DFA} - n_{\sys_R})]$
is the more general equivalent.

\subsection{Using energy decomposition to assess densities}
\label{sec:Decomposition}

Kim et al~\cite{Kim2013-DensDriven} proposed a rather more
practical way of assessing densities, that has important
consequences for functional development.
They highlighted that the DFA error,
$E - E^{\DFA}$ may be decomposed into two more
useful sources of error
-- which they called functional-driven, $\Delta E_F$ and
density-driven, $\Delta E_D$, errors.
These are defined to take the form,
\begin{align}
\Delta E_{F,\sys}^{\DFA}=&E^{\DFA}[n_{\sys}] - E[n_{\sys}] \;,
\\
\Delta E_{D,\sys}^{\DFA}=&E^{\DFA}[n_{\sys}^{\DFA}] - E^{\DFA}[n_{\sys}]\;,
\end{align}
where the functional-driven error reflects the
the use of a DFA at the exact density;
and the density-driven error reflects the use of
the wrong density within the DFA.
Eqs.~\eqref{eqn:Esys} and \eqref{eqn:EsysDFA} reveal
that,
\begin{align}
\Delta E_{F,\sys}^{\DFA}=&E_{\xc}^{\DFA}[n_{\sys}] - E_{\xc}[n_{\sys}]\;,
\end{align}
is the difference between the exact and
approximate xc energies evaluated at the
exact density.

$\Delta E_{D,\sys}^{\DFA}$ is undeniably a useful practical measure of the quality of
densities -- especially once Hartree-Fock densities are used as a substitute ($n_{\sys}:\approx n_{\sys}^{\text{HF}}$) for exact ones.
However, its value does depend on the specific DFA chosen which makes it non-fundamental.
It is therefore useful to use eqs.~\eqref{eqn:Esys} and \eqref{eqn:EsysDFA} to further decompose the energy as,
\begin{align}
\Delta E_{D,\sys}^{\DFA}=&
\Delta E_{D\xc,\sys}^{\DFA} + \Delta E_{D\MF,\sys}^{\DFA}\;.
\end{align}
That is, $\Delta E_D$ is the sum of a
\DFA-dependent xc error,
\begin{align}
\Delta E_{D\xc,\sys}^{\DFA}=&
E_{\xc}^{\DFA}[n^{\DFA}_{\sys}] - E_{\xc}^{\DFA}[n_{\sys}]\;,
\label{eqn:DExc}
\end{align}
and a DFA-independent (fundamental) mean-field error,
\begin{align}
\Delta E_{D\MF,\sys}^{\DFA}=&
F_{\MF}[n^{\DFA}_{\sys}] - F_{\MF}[n_{\sys}]\;.
\label{eqn:DEMF}
\end{align}

Again, it is useful to avoid systematic errors by
looking at errors in energy differences.
To this end,
\begin{align}
\Delta^2 E_{\ldots,\sys}^{\DFA}=&\Delta E_{\ldots,\sys}
-\sum_{A\in \sys} \Delta E_{\ldots,A}
\label{eqn:D2EMF}
\end{align}
is a useful ``double $\Delta$'' expression that
investigates the change in the mean field error
across a process.
Here, $A$ indicates subsystems (e.g., atoms)
of $\sys$, while $\ldots$ indicates
$F$, $D\MF$ or $D\xc$.

An alternative decomposition, [eq.~(45) of \rcite{Vuckovic2019}]
\begin{align}
\Delta E=&\Delta E_{\xc,\sys}^{\DFA} + \Delta E_{\text{ideal},\sys}
\end{align}
splits the energy error into an xc error,
$\Delta E_{\xc,\sys}=E^{\DFA}[n_{\sys}^{\DFA}] - E[n_{\sys}^{\DFA}]$
and an ``ideal'' density error,
$\Delta E_{\text{ideal},\sys}=E[n_{\sys}^{\DFA}] - E[n_{\sys}]$
The ideal density error is then,
\begin{align}
\Delta E_{\text{ideal},\sys}=&
E_{\xc}[n_{\sys}^{\DFA}] - E_{\xc}[n_{\sys}] + \Delta E_{D\MF,\sys}\;,
\end{align}
which includes the MF error as well as the corresponding
xc term.

\subsection{Mean-field error}
\label{sec:MFError}

The mean-field error of eq.~\eqref{eqn:DEMF} 
forms part of the density-driven error.
Importantly, $\Delta E_{D\MF,\sys}$,
offers a pragmatic way to assess the quality of DFA
densities \emph{entirely} independently from the DFA
energies. $\Delta E_{D\MF}$ has
the following (mostly useful) properties:
\begin{enumerate}
\item It does not contain $E_{\xc}$ or $E_{\xc}^{\DFA}$
and therefore tests only the quality of densities;
\item It is zero when $n^{\DFA}=n$;
\item But it can also be zero for other densities
so is therefore not a true metric;
\item It is related to the true xc potential,
and thus indirectly assesses properties related
thereto; (note, $\Delta E_{D\xc}$ similarly
assesses the approximate xc potential)
\item It allows for cancellation of numerical
and basis set errors in kinetic energies and
nuclear energies;
\item Given access to a ``KS inversion code''
(i.e. an algorithm to determine $v_s[n]$ for
a given $n$) it is easy to compute.
\end{enumerate}
The first two points follow directly from the
definition in eq.~\eqref{eqn:DEMF}.
The third point follows from the fact that $F_{\MF}$
involves linear ($T_s + \int nv_{\sys}(\vr)d\vr$),
and quadratic ($E_{\Hrm}$) functionals of
the density so can have multiple zeros.

The fourth point may be understood by considering
the case that the DFA density is close to the true
density, i.e. $\Delta n^{\DFA}:=n^{\DFA}-n\to 0$.
Eq.~\eqref{eqn:DEMF} then yields,
\begin{align}
\Delta E_{D\MF,\sys}
\approx&\int \frac{\delta F_{\MF}}{\delta n(\vr)}
[ n_{\sys}^{\DFA}(\vr) - n_{\sys}(\vr)] d\vr
\nonumber\\
=&-\int  v_{\xc,\sys}(\vr) \Delta n_{\sys}^{\DFA}(\vr) d\vr\;,
 \label{eqn:TDEMF}
\end{align}
using $\tfrac{\delta }{\delta n}T_s=-v_s
=-v_{\sys}-v_{\Hrm}-v_{\xc}$,
$\tfrac{\delta }{\delta n} \int n v_{\sys}d\vr = v_{\sys}$
and $\tfrac{\delta }{\delta n}E_{\Hrm}=v_{\Hrm}$.
Similarly, manipulation of Eqs~\eqref{eqn:DExc}
yields,
\begin{align}
\Delta E_{D\xc,\sys}
\approx& \int  v_{\xc,\sys}^{\DFA}(\vr)
 \Delta n_{\sys}^{\DFA}(\vr) d\vr\;,
 \label{eqn:TDExc}
\end{align}
involving the approximate xc potential.
Taking the sum of Eqs~\eqref{eqn:TDEMF} and
\eqref{eqn:TDExc} gives an error,
$\int (v_{\xc,\sys}-v_{\xc,\sys}^{\DFA})\Delta n_{\sys}^{\DFA}d\vr$,
which Vuckovic et al~\cite{Vuckovic2019} showed
is second order in the $\Delta n_{\sys}^{\DFA}$,
meaning the exact and approximate potentials
approximately differ (up to a constant) in proportion
to $\Delta n_{\sys}^{\DFA}$.
Finally, Eq.~\eqref{eqn:D2EMF} becomes,
\begin{align}
\Delta^2 E_{D\MF}
\approx&-\int  \bigg(
v_{\xc,\sys}\Delta n_{\sys}^{\DFA}
- \sum_A v_{\xc,A} \Delta n_A^{\DFA}
\bigg) d\vr 
 \label{eqn:TD2EMF}
 \end{align}
as the MF energy difference for small changes to densities.
Eq.~\eqref{eqn:TD2EMF} reveals that systematic
errors compensation between
$v_{\xc,\sys}$ and $v_{\xc,A}$
or $\Delta n^{\DFA}_{\sys}$ and
$\Delta n^{\DFA}_A$ will also cancel out.

The fifth point warrants some additional
discussion. Formally, it follows from the fact
that $|E_{\xc}^{(\DFA)}|\ll |F_{\MF}|$ meaning that
the density that minimizes $F_{\MF}+E_{\xc}^{(\DFA)}$
should either be close to the density that minimizes
$F_{\MF}$, or give a value of $F_{\MF}$ that is
close to the minima.
Therefore, variational principles dictate that
$F_{\MF}$ will be relatively insensitive to numerical
errors in $n_{\sys}$ or $n_{\sys}^{\DFA}$.
To illustrate this in practice, consider the case of
atomic Hydrogen, whose properties
($n=\tfrac{1}{\pi}e^{-2r}$,
$T_s=\half=0.5$, $\int n vd\vr =-1$ and
$E_{\Hrm}=-E_{\xc}=\tfrac{5}{16}=0.3125$,
all in Hartree units) are known exactly.
The orbital ($\phi=e^{-r}/\sqrt{\pi}$) may
be approximated using one-, two- and three-Gaussian
models, with coefficients chosen to minimize the
true energy.
The error in $T_s$ is $-0.076$, $-0.014$
and $-0.003$~Ha, for one-, two- and
three-Gaussian models. But, they are balanced
by errors of the opposite sign in $\int n vd\vr$.
The error in $F_{\MF}$ is therefore substantially
smaller -- being $-0.012$~Ha for one-,
and well under 0.001~Ha for two- and
three-Gaussian models.

The sixth point follows from the fact that, of the
ingredients of $F_{\MF}[n]$, only $T_s[n]$
is not trivially computable given the density.
But, given the exact KS potential, $v_s[n]$, one may
determine the exact KS orbitals, $\varphi_k$,
which obey,
$[-\half\nabla^2 + v_s]\varphi_k=\varepsilon_k\varphi_k$.
Thus, one may obtain the KS kinetic energy,
$T_s[n]=\sum_k f_k\int |\nabla\varphi_k|^2 d\vr$.

Taken together, these qualities yield an \emph{exact}
measure that is reasonably easy to compute,
reasonably insensitive to basis errors, yet provides
a meaningful measure of density quality,
despite not being a true metric.
The next section will explore how well it works in
practice, on a series of selected examples, and with
the goal of illustrating its usefulness on practical
chemical problems.

\section{Results}
\label{sec:Results}

\begin{figure}[t!]
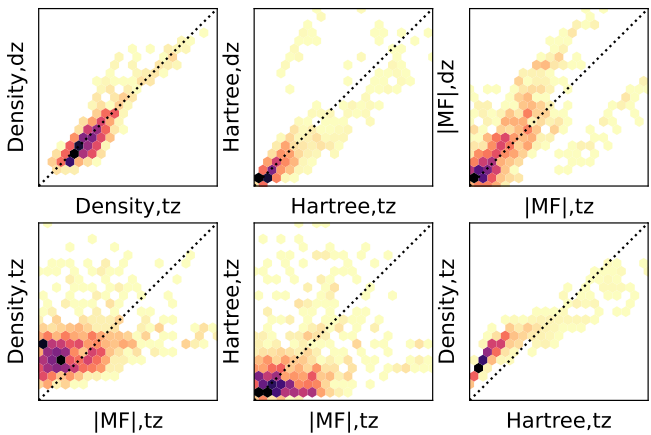

\includegraphics[width=\linewidth]{{{FigCompare}}}
\caption{Comparisons between various error measures
across the full suite of random molecules, with all errors
normalized by their average value.
Here, `Hartree' indicates eq.~\eqref{eqn:ErrH},
`Density' indicates eq.~\eqref{eqn:ErrD} and
`MF' indicates eq.~\eqref{eqn:FMF}.
`dz'/`tz' indicate the cc-pvdz/cc-pvtz basis set.
Dark colours indicate many points, yellow indicates
few points, and white indicates no points.
Dotted lines show $y=x$, so indicate perfect
linear correlation.
\label{fig:Compare} }
\end{figure}

All computation in this section is done
using customized {\tt psi4}-based code.%
~\cite{psi4,psi4numpy}
Most calculations use moderate (cc-pvtz~\cite{Weigend2005}, tz)
quality basis, although sometimes lower
(cc-pvdz, dz) quality sets are used to test
the impact of basis set on results.

$F_{\MF}[n_{\sys}^{\DFA}]$ is
computed as part of any DFT calculation.
For the work here orbitals are obtained from
KS theory (restricted for even and unrestricted for
odd numbers of electrons) and used directly to evaluate
$T_s$ and $\int n v_{\sys}d\vr$. $E_{\Hrm}$ is
computed using density-fitted electron repulsion
integrals (DF-ERIs).~\cite{Weigend2005}
This ensures helps to improve consistency with
the reference results, which use DF-ERIs as
part of the inversion procedure.

Reference mean-field energies,
$F_{\MF}[n_{\sys}]$, are obtained from
coupled cluster singles and doubles (CCSD)
densities. The resulting density-matrix can be used
to directly evaluate
$\int n v_{\sys}d\vr$ and $E_{\Hrm}$ in the
same way as above.
Reference values for $T_s$ are obtained using the
method described in \rcite{Gould2023}, also with
the same DF-ERI basis used to evaluate $E_{\Hrm}$.
For systems with odd numbers of electrons
reference values are at the restricted-orbital
Kohn-Sham level, where $\up$ and $\down$
electrons share the same spatial orbitals.

Results are reported for (up to) fifteen DFAs
covering multiple rungs of Jacob's ladder;~\cite{Perdew2001-Jacob}
and with a mix of minimally
empirical (where only a few `free' parameters are
optimized over energies) and empirical (where most
parameters are optimized over energies)
construction. The investigated DFAs are:
\begin{itemize}
\item The local density approximation (LDA)
using VWN~\cite{g09_svwn} correlation;
\item The minimally empirical GGAs PBE~\cite{g09_pbepbe},
BLYP~\cite{g09_blyp},
and empirical N12~\cite{g09_N12} and SOGGA11~\cite{g09_SOGGA11};
\item The minimally empirical meta-GGAs
TPSS~\cite{g09_tpsstpss} and SCAN~\cite{DFA:SCAN};
and more empirical M06-L~\cite{g09_M06L}
and MN15-L~\cite{g09_MN15L};
\item Hartree-Fock theory; the minimally empirical
hybrids GGAs PBE0~\cite{g09_pbe1pbe} and B3LYP~\cite{g09_b3lyp};
and the more empirical M06~\cite{g09_m06};
\item The empirical range-separated hybrid,
$\omega$B97X-V~\cite{g09_wB97x},
and the minimally empirical double hybrid
B2PLYP~\cite{g09_b2plyp} (implemented in the usual
way with MP2 contributions to the energies,
but not to the densities).
\end{itemize}

\subsection{C and H systems}

The first test of density-related properties involves
56 random molecules and ions containing
only C and H -- 17 CH$_3$ anions,
19 neutral CH$_4$, and 20 CH$_5$ cations.%
~\footnote{Twenty of each were initially computed,
but cases where one or more DFA did not
converge where excluded from the final list.}
The structures are random, but obey the
following rules: 1) CH distances are between
1 and 1.8~{\AA} at 0.1~{\AA} steps;
2) angles are random but all HH distances must
be greater than 0.7~\AA.
Using these random systems controls for
elemental effects on energies (since only C and
H are involved) but provides a
reasonable diversity of chemical interactions.

To begin, consider the usefulness of
$E_{D\MF}$. Figure~\ref{fig:Compare} compares
different measures of density errors, for dz and tz
basis sets, for all 56 molecules and for all 15 DFAs.
The top row of plots compares results for
tz and dz, to reveal that, despite the low
quality of a dz basis, most errors
correlate rather closely in both basis sets
-- although there are some exceptions.
tz is thus chosen for subsequent work.

The bottom row indicates that Hartree and
density errors are closely correlated. But,
it also reveals something rather alarming.
There are many systems with large
mean field errors but small density
or Hartree errors. That is, there are many
systems where the DFA density has
a more substantial impact on the
DFA energy than is captured by common
metrics. This suggests that typical metrics may
miss important features of densities.

\begin{figure}[h!]
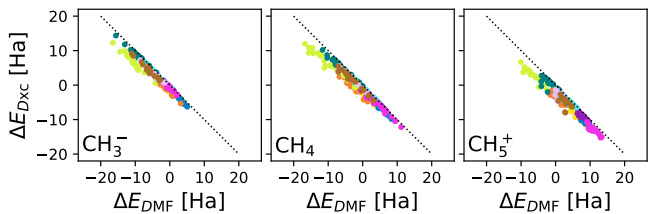

\includegraphics[width=\linewidth]{{{FigEnergy}}}
\caption{Deviations of xc and mean-field energies
for different DFAs.
Colours identify DFAs and are the same as in
Figure~\ref{fig:DFAs}.
Dotted lines indicate $y=-x$ and thus perfect
cancellation of errors.
\label{fig:Energy} }
\end{figure}

Next, consider the total density-driven error.
Figure~\ref{fig:Energy} shows the MF and xc parts
of this error, which cancel almost perfectly in all
cases, as expected and shown in \rcite{Vuckovic2019}.
As shall become apparent later, this nearly perfect
cancellation is not univeral. Most likely, it
at least partly reflects the fact that DFAs become
popular because they work on organic chemistry
-- and especially on CH bonds.
The exceptional case is MN15-L (in yellow) which has
noticeably less cancellation of errors than other
DFAs.

\begin{figure}[h!]
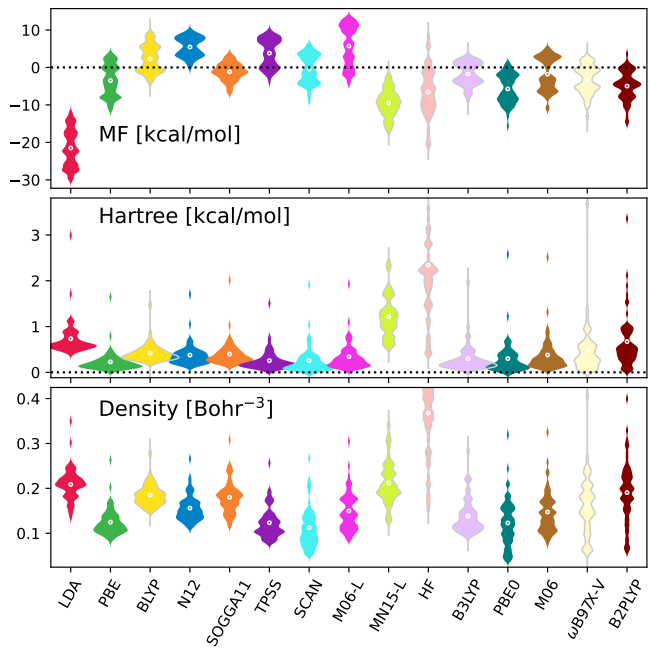

\includegraphics[width=\linewidth]{{{FigDFAs}}}
\caption{Distribution of different density errors
for selected DFAs.
Note, Hartree and density errors below zero
are an artefact of the violin plot model.
\label{fig:DFAs} }
\end{figure}

Finally, consider how $\Delta E_{D\MF}$
varies with different DFAs, and how this compares
with more traditional metrics for density errors.
Figure~\ref{fig:DFAs} shows the error distribution
(as violin plots whose width indicates the
probability of errors)
over all 56 molecules for each of the 15 considered
DFAs. DFAs are grouped according to
rung/complexity.

The most obvious conclusion to be drawn is that
most functionals yield very low Hartree errors --
less than 1~kcal/mol in the vast majority of cases.
Variation in density errors is a little greater, but
is broadly consistent with the Hartree errors.
Surprisingly, Hartree-Fock (HF) and B2PLYP (with
a high fraction of Hartree-Fock) densities are
rather poor -- despite B2PLYP giving excellent
energies. This supports the cautionary note in
recent work~\cite{Sim2022} on using
density-corrected DFT (i.e. using HF densities and orbitals
to compute DFA energies) as a general purpose
tool.

The MF error provides a more complex picture of DFAs.
Firstly, it reveals that the best MF errors are from
SOGGA11, SCAN and B3LYP, which
give $\Delta E_{D\MF}$ within $\pm 5$~kcal/mol
in the majority of cases.
Notably, SCAN outperforms PBE0 despite not using
any exact Hartree-Fock exchange, perhaps reflecting
its satisfaction of many exact limits.~\cite{DFA:SCAN}
Secondly, Hartree-Fock is the least consistent
performer -- all DFAs have a narrower distribution
of errors which suggests that inclusion of even
approximate correlation generally improves densities.
Thirdly, of the remaining heavily parameterized DFAs
N12, M06-L, MN15L, M06 and $\omega$B97X-V,
only M06 and $\omega$B97X-V (both hybrid DFAs)
do a good job on densities across all three errors.
It is also notable that SOGGA11 is also not as good
as SCAN or B3LYP on the other errors, despite its
success on $\Delta E_{D\MF}$.

Finally, Figure~\ref{fig:DFAs} reveals an interesting
result pertaining to the ``path to exactness''.
LDA, PBE, TPSS and SCAN (all except
LDA from Perdew and co-workers) are clearly
\emph{staying on} the path to exactness.
Each step leads to satisfaction of more exact
constraints, and each step yields
an improvement to the mean-field error.

\subsection{Cl--OH$^-$ system}

\begin{figure}[h!]
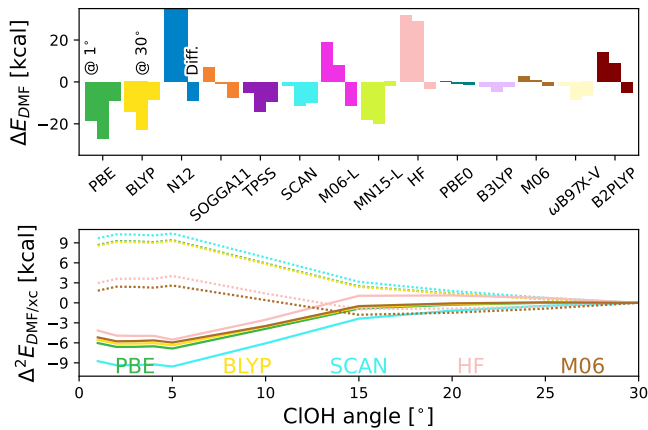

\includegraphics[width=\linewidth]{{{FigHOClm}}}
\caption{
Mean-field and density-driven errors
for HO-Cl$^-$ with various HOCl angles.
Mean-field errors (top) are shown for $1^{\circ}$ and
25$^{\circ}$, as well as the difference between the two.
Both components of density-driven errors (bottom)
are shown for all angles, for a more limited set of DFAs
-- xc errors are solid lines and MF errors are dotted lines.
$0^{\circ}$ is excluded due to issues in treating
the higher-symmetry structure. LDA is excluded
as it did not converge for angles $<5^{\circ}$.
\label{fig:HOClm} }
\end{figure}

Having demonstrated the usefulness of the mean-field
error, let us now turn to problems where densities
can be a genuine problem.
The OH--Cl$^-$ dimer has been shown to benefit
from density-corrected (DC-)DFT, in which
HF densities are used together with
standard DFAs for energies to avoid non-systematic
cancellation of errors as ClOH angles are increased.
By using DC-DFT, the zero angle structure is
(correctly) predicted to be optimal, versus 25$^{\circ}$
or so for self-consistent DFAs.~\cite{Sim2022}
This success has been justified by the fact that
HF densities avoid the delocalization errors of GGAs.

Figure~\ref{fig:HOClm} illustrates that DC-DFT
also benefits from cancellation of errors. In fact,
as the top plot shows, the HF densities are quite
bad according to $\Delta E_{D\MF}$ -- only N12 is worse.
But, the HF error increases only slightly as angles
are increased, whereas all GGAs and all meta-GGAs
bar MN15-L predict a much larger increase.
The xc contribution to the density-driven error
(bottom plot) is much more consistent across
different DFAs, which means that using the
HF density can lead to signficant changes in
energy ordering.

\subsection{He$_2^+$ system}

\begin{figure}[h!]
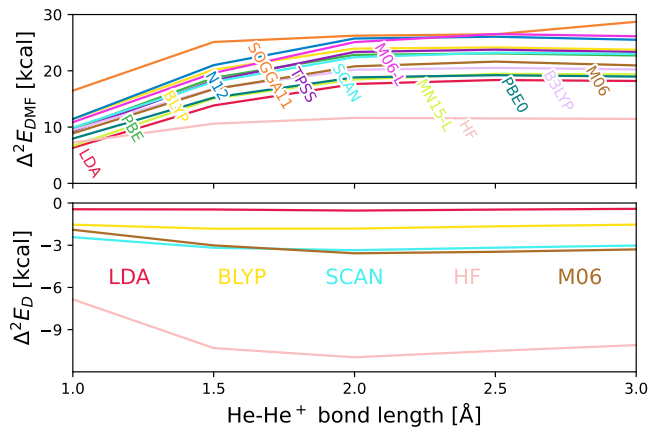

\includegraphics[width=\linewidth]{{{FigHe2p}}}
\caption{
Mean-field (top) and density-driven (bottom) errors
for He$_2^+$ at various bond lengths.
\label{fig:He2p} }
\end{figure}

The last example to be investigated is He$_2^+$,
which represents a stringent test of densities as
the spare electron is shared equally between the
two cations when they are close together, but
represents a difficult dissociation into
He$+$He$^+=2\times$He$^{+0.5}$ when they are pulled apart.
Figure~\ref{fig:He2p} shows mean-field and
density-driven errors relative to the error in
He and He$^+$ -- which should be (but isn't)
the limit of large bond length.

Previous work~\cite{Song2021} has shown that
density-driven errors do not have much effect on this
system, which is confirmed by this work.
Nevertheless, HF is the standout performer for densities,
with systematic mean field errors of -10~kcal/mol.
By contrast, all DFAs start with similar errors to
HF, but get worse as the molecule as dissociated.
The total density-driven error reveals that the
mean-field errors are almost entirely cancelled by
the corresponding density-driven xc errors -- with
the \emph{exception} of HF theory. Breaking
apart the density-driven error into its mean-field
and xc components therefore provides a more nuanced
picture of what is going on in this difficult
exemplar case.

\section{The future of density benchmarking}
\label{sec:Future}

Due to their direct connection to energies,
$\Delta E_{D\MF}$ and $\Delta^2 E_{D\MF}$
are promising tools for analysing errors in densities
independent of any approximations
As can be seen in the previous section, both
offer insights that aren't available in traditional
analysis and assessment of densities.
The increasing quality and reliability of
general purpose Kohn-Sham inversion algorithms,%
~\cite{Wang1993,Wu2003,Kanungo2019,Shi2021,Gould2023}
and relative insensitivity of $\Delta E_{D\MF}$ to
basis set effects make future routine use quite feasible.
Both measures thus provide a crucial first step to
systematic benchmarking of densities.
The author outlines several promising directions,
and remaining challenges, below.

{\bf Assessment of DFAs:}
The results from Sec~\ref{sec:Results}
illustrate that all DFAs benefit
from cancellation of density- and energy-based
errors, as can be expected (but not guaranteed)
from minimization principles.
For the systems tested here, only N12
and HF have consistently poor densities, as
shown by large systematic errors (N12) or
highly variable errors (HF).
Results also suggest that the sequence of
LDA, PBE, TPSS, SCAN is staying on the exact path.
A deeper analysis involving a wider range of
exemplar chemical physics would be
required to properly assess whether highly
empirical DFAs are truly straying from the path
to exactness, or merely taking a side road.

{\bf DFAs for densities:}
The work suggests that DFAs might be
optimized to improve densities, at the expense
of energies. Factoring densities into
optimization strategies may offer improvements
on DC-DFT~\cite{Sim2022}, which already
improves on self-consistent calculations when
used sensibly.
For example, despite being optimized on
energies, B3LYP clearly out-performs its
``ingredients'' (LDA, BLYP and HF) on
densities. Rigorous optimization of DFAs
to improve densities may offer a ``best of
both worlds'' route to DFAs, in which
densities are treated accurately by one
approach and energies by another.

{\bf Benchmarking the benchmarks:}
One important aspect that was brushed
over in the previous section is the
reliability of benchmark data.
The author has followed tradition in using
CCSD densities as a reference. But this
begs several questions:
How accurate are CCSD densities?
When do they become inaccurate?
How do we assess inaccuracies?
We need to benchmark the benchmarks to
answer these questions.
[Note, Kanungo et al~\cite{Kanungo2023}
make the case that Brueckner orbital based
coupled cluster theory may be more
appropriate.]

Similarly, the author has assumed that
the cc-pvtz basis set is sufficiently accurate for
benchmarking densities.
For present illustrative purposes this is almost
certainly the case -- indeed, cc-pvdz was
probably sufficient. But, benchmark energies
have benefited from substational
development and refinement of techniques
to extrapolate to the basis set limit, especially
via composite protocols (e.g. Weizmann-4
theory~\cite{Karton2006}).
Can (and should?) something similar be done
for densities?

\section{Conclusions}

DFAs serve a major role in computational
electronic structure theory. Addressing the growing
need for accurate DFAs has pushed human-optimized
DFAs to their limits -- SCAN~\cite{DFA:SCAN} obeys an unprecedented
17 constraints. Empiricism and data-driven
(machine learned) DFAs can fill in gaps.
But, data-driven approaches can optimize for
cancellation of errors in the training set (interpolation),
which can lead to problems when they are
applied to novel chemical physics (extrapolation).
There is thus an urgent need to use more
physical constraints in their construction.
Otherwise, there is a risk of developing DFAs
that have outstanding interpolative abilities,
but are poorer than the current state-of-art
at extrapolation -- which is precisely where
they are needed.

Testing, improving and even optimizing on
the quality of densities is a promising way to
incorporate additional physical data into DFAs.
The mean-field error, $\Delta E_{D\MF}$,
[eq.~\eqref{eqn:DEMF}] presented here
offers several advantages in this regard,
as detailed in Section~\ref{sec:MFError}.
Its usefulness as a measure is illustrated in this
work using selected examples in Section~\ref{sec:Results}.

The present work presents multiple future
challenges and directions [see also
Section~\ref{sec:Future}].
Firstly, new benchmarking strategies need
to be developed to generate useful benchmark
data for densities. Secondly, those strategies
need to be turned into useful high-quality
reference data that can be used to test and develop DFAs.
Finally, a more diverse range of systems needs
to be studied, to offer insights into which DFAs
do better, and worse, at predicting densities.

Once these challenges have been met, the resulting
benchmark sets of $\Delta F_{\MF}$ will
allow data-driven DFA to incorporate densities into
their optimization, and so ensure they stay on the path
to exactness, by design.

\acknowledgements
TG thanks Amir Karton for helpful discussion on
composite benchmarking protocols, and Stefan
Vuckovic for highlighting additional connections to
the theory of density-corrected DFT.
TG was supported by an Australian Research Council (ARC)
Discovery Project (DP200100033) and Future Fellowship (FT210100663).

\bibliography{MFError,DFAList}

\begin{thebibliography}{49}%
\makeatletter
\providecommand \@ifxundefined [1]{%
 \@ifx{#1\undefined}
}%
\providecommand \@ifnum [1]{%
 \ifnum #1\expandafter \@firstoftwo
 \else \expandafter \@secondoftwo
 \fi
}%
\providecommand \@ifx [1]{%
 \ifx #1\expandafter \@firstoftwo
 \else \expandafter \@secondoftwo
 \fi
}%
\providecommand \natexlab [1]{#1}%
\providecommand \enquote  [1]{``#1''}%
\providecommand \bibnamefont  [1]{#1}%
\providecommand \bibfnamefont [1]{#1}%
\providecommand \citenamefont [1]{#1}%
\providecommand \href@noop [0]{\@secondoftwo}%
\providecommand \href [0]{\begingroup \@sanitize@url \@href}%
\providecommand \@href[1]{\@@startlink{#1}\@@href}%
\providecommand \@@href[1]{\endgroup#1\@@endlink}%
\providecommand \@sanitize@url [0]{\catcode `\\12\catcode `\$12\catcode
  `\&12\catcode `\#12\catcode `\^12\catcode `\_12\catcode `\%12\relax}%
\providecommand \@@startlink[1]{}%
\providecommand \@@endlink[0]{}%
\providecommand \url  [0]{\begingroup\@sanitize@url \@url }%
\providecommand \@url [1]{\endgroup\@href {#1}{\urlprefix }}%
\providecommand \urlprefix  [0]{URL }%
\providecommand \Eprint [0]{\href }%
\providecommand \doibase [0]{https://doi.org/}%
\providecommand \selectlanguage [0]{\@gobble}%
\providecommand \bibinfo  [0]{\@secondoftwo}%
\providecommand \bibfield  [0]{\@secondoftwo}%
\providecommand \translation [1]{[#1]}%
\providecommand \BibitemOpen [0]{}%
\providecommand \bibitemStop [0]{}%
\providecommand \bibitemNoStop [0]{.\EOS\space}%
\providecommand \EOS [0]{\spacefactor3000\relax}%
\providecommand \BibitemShut  [1]{\csname bibitem#1\endcsname}%
\let\auto@bib@innerbib\@empty
\bibitem [{\citenamefont {Hohenberg}\ and\ \citenamefont
  {Kohn}(1964)}]{HohenbergKohn}%
  \BibitemOpen
  \bibfield  {author} {\bibinfo {author} {\bibfnamefont {P.}~\bibnamefont
  {Hohenberg}}\ and\ \bibinfo {author} {\bibfnamefont {W.}~\bibnamefont
  {Kohn}},\ }\bibfield  {title} {\enquote {\bibinfo {title} {Inhomogeneous
  electron gas},}\ }\href@noop {} {\bibfield  {journal} {\bibinfo  {journal}
  {Phys. Rev.}\ }\textbf {\bibinfo {volume} {136}},\ \bibinfo {pages}
  {B864--B871} (\bibinfo {year} {1964})}\BibitemShut {NoStop}%
\bibitem [{\citenamefont {Kohn}\ and\ \citenamefont {Sham}(1965)}]{KohnSham}%
  \BibitemOpen
  \bibfield  {author} {\bibinfo {author} {\bibfnamefont {W.}~\bibnamefont
  {Kohn}}\ and\ \bibinfo {author} {\bibfnamefont {L.~J.}\ \bibnamefont
  {Sham}},\ }\bibfield  {title} {\enquote {\bibinfo {title} {Self-consistent
  equations including exchange and correlation effects},}\ }\href@noop {}
  {\bibfield  {journal} {\bibinfo  {journal} {Phys. Rev.}\ }\textbf {\bibinfo
  {volume} {140}},\ \bibinfo {pages} {A1133--A1138} (\bibinfo {year}
  {1965})}\BibitemShut {NoStop}%
\bibitem [{\citenamefont {Karton}, \citenamefont {Daon},\ and\ \citenamefont
  {Martin}(2011)}]{Karton2011}%
  \BibitemOpen
  \bibfield  {author} {\bibinfo {author} {\bibfnamefont {A.}~\bibnamefont
  {Karton}}, \bibinfo {author} {\bibfnamefont {S.}~\bibnamefont {Daon}},\ and\
  \bibinfo {author} {\bibfnamefont {J.~M.}\ \bibnamefont {Martin}},\ }\bibfield
   {title} {\enquote {\bibinfo {title} {W4-11: A high-confidence benchmark
  dataset for computational thermochemistry derived from first-principles w4
  data},}\ }\href@noop {} {\bibfield  {journal} {\bibinfo  {journal} {Chem.
  Phys. Lett.}\ }\textbf {\bibinfo {volume} {510}},\ \bibinfo {pages}
  {165--178} (\bibinfo {year} {2011})}\BibitemShut {NoStop}%
\bibitem [{\citenamefont {Mardirossian}\ and\ \citenamefont
  {Head-Gordon}(2017)}]{Mardirossian2017}%
  \BibitemOpen
  \bibfield  {author} {\bibinfo {author} {\bibfnamefont {N.}~\bibnamefont
  {Mardirossian}}\ and\ \bibinfo {author} {\bibfnamefont {M.}~\bibnamefont
  {Head-Gordon}},\ }\bibfield  {title} {\enquote {\bibinfo {title} {Thirty
  years of density functional theory in computational chemistry: an overview
  and extensive assessment of 200 density functionals},}\ }\href@noop {}
  {\bibfield  {journal} {\bibinfo  {journal} {Mol. Phys.}\ }\textbf {\bibinfo
  {volume} {115}},\ \bibinfo {pages} {2315--2372} (\bibinfo {year}
  {2017})}\BibitemShut {NoStop}%
\bibitem [{\citenamefont {Goerigk}\ \emph {et~al.}(2017)\citenamefont
  {Goerigk}, \citenamefont {Hansen}, \citenamefont {Bauer}, \citenamefont
  {Ehrlich}, \citenamefont {Najibi},\ and\ \citenamefont
  {Grimme}}]{Goerigk2017}%
  \BibitemOpen
  \bibfield  {author} {\bibinfo {author} {\bibfnamefont {L.}~\bibnamefont
  {Goerigk}}, \bibinfo {author} {\bibfnamefont {A.}~\bibnamefont {Hansen}},
  \bibinfo {author} {\bibfnamefont {C.}~\bibnamefont {Bauer}}, \bibinfo
  {author} {\bibfnamefont {S.}~\bibnamefont {Ehrlich}}, \bibinfo {author}
  {\bibfnamefont {A.}~\bibnamefont {Najibi}},\ and\ \bibinfo {author}
  {\bibfnamefont {S.}~\bibnamefont {Grimme}},\ }\bibfield  {title} {\enquote
  {\bibinfo {title} {A look at the density functional theory zoo with the
  advanced {GMTKN55} database for general main group thermochemistry, kinetics
  and noncovalent interactions},}\ }\href@noop {} {\bibfield  {journal}
  {\bibinfo  {journal} {Phys. Chem. Chem. Phys.}\ }\textbf {\bibinfo {volume}
  {19}},\ \bibinfo {pages} {32184--32215} (\bibinfo {year} {2017})}\BibitemShut
  {NoStop}%
\bibitem [{\citenamefont {Chan}, \citenamefont {Gill},\ and\ \citenamefont
  {Kimura}(2019)}]{Chan2019}%
  \BibitemOpen
  \bibfield  {author} {\bibinfo {author} {\bibfnamefont {B.}~\bibnamefont
  {Chan}}, \bibinfo {author} {\bibfnamefont {P.~M.~W.}\ \bibnamefont {Gill}},\
  and\ \bibinfo {author} {\bibfnamefont {M.}~\bibnamefont {Kimura}},\
  }\bibfield  {title} {\enquote {\bibinfo {title} {Assessment of {DFT} methods
  for transition metals with the {TMC}151 compilation of data sets and
  comparison with accuracies for main-group chemistry},}\ }\href
  {https://doi.org/10.1021/acs.jctc.9b00239} {\bibfield  {journal} {\bibinfo
  {journal} {J. Chem. Theory Comput.}\ }\textbf {\bibinfo {volume} {15}},\
  \bibinfo {pages} {3610--3622} (\bibinfo {year} {2019})}\BibitemShut {NoStop}%
\bibitem [{\citenamefont {Tawfik}\ \emph {et~al.}(2018)\citenamefont {Tawfik},
  \citenamefont {Gould}, \citenamefont {Stampfl},\ and\ \citenamefont
  {Ford}}]{Tawfik2018}%
  \BibitemOpen
  \bibfield  {author} {\bibinfo {author} {\bibfnamefont {S.~A.}\ \bibnamefont
  {Tawfik}}, \bibinfo {author} {\bibfnamefont {T.}~\bibnamefont {Gould}},
  \bibinfo {author} {\bibfnamefont {C.}~\bibnamefont {Stampfl}},\ and\ \bibinfo
  {author} {\bibfnamefont {M.~J.}\ \bibnamefont {Ford}},\ }\bibfield  {title}
  {\enquote {\bibinfo {title} {Evaluation of van der waals density functionals
  for layered materials},}\ }\href
  {https://doi.org/10.1103/physrevmaterials.2.034005} {\bibfield  {journal}
  {\bibinfo  {journal} {Phys. Rev. Materials}\ }\textbf {\bibinfo {volume}
  {2}},\ \bibinfo {pages} {034005} (\bibinfo {year} {2018})}\BibitemShut
  {NoStop}%
\bibitem [{\citenamefont {Seeger}\ and\ \citenamefont
  {Izgorodina}(2020)}]{Seeger2020}%
  \BibitemOpen
  \bibfield  {author} {\bibinfo {author} {\bibfnamefont {Z.~L.}\ \bibnamefont
  {Seeger}}\ and\ \bibinfo {author} {\bibfnamefont {E.~I.}\ \bibnamefont
  {Izgorodina}},\ }\bibfield  {title} {\enquote {\bibinfo {title} {A systematic
  study of {DFT} performance for geometry optimizations of ionic liquid
  clusters},}\ }\href {https://doi.org/10.1021/acs.jctc.0c00549} {\bibfield
  {journal} {\bibinfo  {journal} {J. Chem. Theory Comput.}\ }\textbf {\bibinfo
  {volume} {16}},\ \bibinfo {pages} {6735--6753} (\bibinfo {year}
  {2020})}\BibitemShut {NoStop}%
\bibitem [{\citenamefont {Mester}\ and\ \citenamefont
  {K{\'{a}}llay}(2022)}]{Mester2022}%
  \BibitemOpen
  \bibfield  {author} {\bibinfo {author} {\bibfnamefont {D.}~\bibnamefont
  {Mester}}\ and\ \bibinfo {author} {\bibfnamefont {M.}~\bibnamefont
  {K{\'{a}}llay}},\ }\bibfield  {title} {\enquote {\bibinfo {title}
  {Charge-transfer excitations within density functional theory: How accurate
  are the most recommended approaches?}}\ }\href
  {https://doi.org/10.1021/acs.jctc.1c01307} {\bibfield  {journal} {\bibinfo
  {journal} {J. Chem. Theory Comput.}\ }\textbf {\bibinfo {volume} {18}},\
  \bibinfo {pages} {1646--1662} (\bibinfo {year} {2022})}\BibitemShut {NoStop}%
\bibitem [{\citenamefont {Perdew}\ and\ \citenamefont
  {Schmidt}(2001)}]{Perdew2001-Jacob}%
  \BibitemOpen
  \bibfield  {author} {\bibinfo {author} {\bibfnamefont {J.~P.}\ \bibnamefont
  {Perdew}}\ and\ \bibinfo {author} {\bibfnamefont {K.}~\bibnamefont
  {Schmidt}},\ }\bibfield  {title} {\enquote {\bibinfo {title}
  {Jacob{\textquoteright}s ladder of density functional approximations for the
  exchange-correlation energy},}\ }\href@noop {} {\bibfield  {journal}
  {\bibinfo  {journal} {AIP Conf. Proc.}\ }\textbf {\bibinfo {volume} {577}},\
  \bibinfo {pages} {1--20} (\bibinfo {year} {2001})}\BibitemShut {NoStop}%
\bibitem [{\citenamefont {Gould}\ and\ \citenamefont
  {Dale}(2022)}]{Gould2022-Poison}%
  \BibitemOpen
  \bibfield  {author} {\bibinfo {author} {\bibfnamefont {T.}~\bibnamefont
  {Gould}}\ and\ \bibinfo {author} {\bibfnamefont {S.~G.}\ \bibnamefont
  {Dale}},\ }\bibfield  {title} {\enquote {\bibinfo {title} {Poisoning density
  functional theory with benchmark sets of difficult systems},}\ }\href
  {https://doi.org/10.1039/d2cp00268j} {\bibfield  {journal} {\bibinfo
  {journal} {Phys. Chem. Chem. Phys.}\ }\textbf {\bibinfo {volume} {24}},\
  \bibinfo {pages} {6398--6403} (\bibinfo {year} {2022})}\BibitemShut {NoStop}%
\bibitem [{\citenamefont {Medvedev}\ \emph {et~al.}(2017)\citenamefont
  {Medvedev}, \citenamefont {Bushmarinov}, \citenamefont {Sun}, \citenamefont
  {Perdew},\ and\ \citenamefont {Lyssenko}}]{Medvedev2017}%
  \BibitemOpen
  \bibfield  {author} {\bibinfo {author} {\bibfnamefont {M.~G.}\ \bibnamefont
  {Medvedev}}, \bibinfo {author} {\bibfnamefont {I.~S.}\ \bibnamefont
  {Bushmarinov}}, \bibinfo {author} {\bibfnamefont {J.}~\bibnamefont {Sun}},
  \bibinfo {author} {\bibfnamefont {J.~P.}\ \bibnamefont {Perdew}},\ and\
  \bibinfo {author} {\bibfnamefont {K.~A.}\ \bibnamefont {Lyssenko}},\
  }\bibfield  {title} {\enquote {\bibinfo {title} {Density functional theory is
  straying from the path toward the exact functional},}\ }\href@noop {}
  {\bibfield  {journal} {\bibinfo  {journal} {Sci}\ }\textbf {\bibinfo {volume}
  {355}},\ \bibinfo {pages} {49--52} (\bibinfo {year} {2017})}\BibitemShut
  {NoStop}%
\bibitem [{\citenamefont {Kepp}(2017)}]{Kepp2017}%
  \BibitemOpen
  \bibfield  {author} {\bibinfo {author} {\bibfnamefont {K.~P.}\ \bibnamefont
  {Kepp}},\ }\bibfield  {title} {\enquote {\bibinfo {title} {Comment on
  {\textquotedblleft}density functional theory is straying from the path toward
  the exact functional{\textquotedblright}},}\ }\href
  {https://doi.org/10.1126/science.aam9364} {\bibfield  {journal} {\bibinfo
  {journal} {Sci}\ }\textbf {\bibinfo {volume} {356}},\ \bibinfo {pages}
  {496--496} (\bibinfo {year} {2017})}\BibitemShut {NoStop}%
\bibitem [{\citenamefont {Gould}(2017)}]{Gould2017-Fukui}%
  \BibitemOpen
  \bibfield  {author} {\bibinfo {author} {\bibfnamefont {T.}~\bibnamefont
  {Gould}},\ }\bibfield  {title} {\enquote {\bibinfo {title} {What makes a
  density functional approximation good? insights from the left fukui
  function},}\ }\href {https://doi.org/10.1021/acs.jctc.7b00231} {\bibfield
  {journal} {\bibinfo  {journal} {J. Chem. Theory Comput.}\ }\textbf {\bibinfo
  {volume} {13}},\ \bibinfo {pages} {2373--2377} (\bibinfo {year}
  {2017})}\BibitemShut {NoStop}%
\bibitem [{\citenamefont {Mezei}, \citenamefont {Csonka},\ and\ \citenamefont
  {K{\'{a}}llay}(2017)}]{Mezei2017}%
  \BibitemOpen
  \bibfield  {author} {\bibinfo {author} {\bibfnamefont {P.~D.}\ \bibnamefont
  {Mezei}}, \bibinfo {author} {\bibfnamefont {G.~I.}\ \bibnamefont {Csonka}},\
  and\ \bibinfo {author} {\bibfnamefont {M.}~\bibnamefont {K{\'{a}}llay}},\
  }\bibfield  {title} {\enquote {\bibinfo {title} {Electron density errors and
  density-driven exchange-correlation energy errors in approximate density
  functional calculations},}\ }\href {https://doi.org/10.1021/acs.jctc.7b00550}
  {\bibfield  {journal} {\bibinfo  {journal} {J. Chem. Theory Comput.}\
  }\textbf {\bibinfo {volume} {13}},\ \bibinfo {pages} {4753--4764} (\bibinfo
  {year} {2017})}\BibitemShut {NoStop}%
\bibitem [{\citenamefont {Hait}\ and\ \citenamefont
  {Head-Gordon}(2018)}]{Hait2018}%
  \BibitemOpen
  \bibfield  {author} {\bibinfo {author} {\bibfnamefont {D.}~\bibnamefont
  {Hait}}\ and\ \bibinfo {author} {\bibfnamefont {M.}~\bibnamefont
  {Head-Gordon}},\ }\bibfield  {title} {\enquote {\bibinfo {title} {How
  accurate is density functional theory at predicting dipole moments? an
  assessment using a new database of 200 benchmark values},}\ }\href
  {https://doi.org/10.1021/acs.jctc.7b01252} {\bibfield  {journal} {\bibinfo
  {journal} {J. Chem. Theory Comput.}\ }\textbf {\bibinfo {volume} {14}},\
  \bibinfo {pages} {1969--1981} (\bibinfo {year} {2018})}\BibitemShut {NoStop}%
\bibitem [{\citenamefont {Br{\'{e}}mond}\ \emph {et~al.}(2021)\citenamefont
  {Br{\'{e}}mond}, \citenamefont {Tognetti}, \citenamefont {Chermette},
  \citenamefont {Sancho-Garc{\'{\i}}a}, \citenamefont {Joubert},\ and\
  \citenamefont {Adamo}}]{Bremond2021}%
  \BibitemOpen
  \bibfield  {author} {\bibinfo {author} {\bibfnamefont {{\'{E}}.}~\bibnamefont
  {Br{\'{e}}mond}}, \bibinfo {author} {\bibfnamefont {V.}~\bibnamefont
  {Tognetti}}, \bibinfo {author} {\bibfnamefont {H.}~\bibnamefont {Chermette}},
  \bibinfo {author} {\bibfnamefont {J.~C.}\ \bibnamefont
  {Sancho-Garc{\'{\i}}a}}, \bibinfo {author} {\bibfnamefont {L.}~\bibnamefont
  {Joubert}},\ and\ \bibinfo {author} {\bibfnamefont {C.}~\bibnamefont
  {Adamo}},\ }\bibfield  {title} {\enquote {\bibinfo {title} {Electronic energy
  and local property errors at {QTAIM} critical points while climbing perdew's
  ladder of density-functional approximations},}\ }\href
  {https://doi.org/10.1021/acs.jctc.1c00981} {\bibfield  {journal} {\bibinfo
  {journal} {J. Chem. Theory Comput.}\ }\textbf {\bibinfo {volume} {18}},\
  \bibinfo {pages} {293--308} (\bibinfo {year} {2021})}\BibitemShut {NoStop}%
\bibitem [{\citenamefont {Landeros-Rivera}\ \emph {et~al.}(2022)\citenamefont
  {Landeros-Rivera}, \citenamefont {Gallegos}, \citenamefont {Mun{\'{a}}rriz},
  \citenamefont {Laplaza},\ and\ \citenamefont
  {Contreras-Garc{\'{\i}}a}}]{LanderosRivera2022}%
  \BibitemOpen
  \bibfield  {author} {\bibinfo {author} {\bibfnamefont {B.}~\bibnamefont
  {Landeros-Rivera}}, \bibinfo {author} {\bibfnamefont {M.}~\bibnamefont
  {Gallegos}}, \bibinfo {author} {\bibfnamefont {J.}~\bibnamefont
  {Mun{\'{a}}rriz}}, \bibinfo {author} {\bibfnamefont {R.}~\bibnamefont
  {Laplaza}},\ and\ \bibinfo {author} {\bibfnamefont {J.}~\bibnamefont
  {Contreras-Garc{\'{\i}}a}},\ }\bibfield  {title} {\enquote {\bibinfo {title}
  {New venues in electron density analysis},}\ }\href
  {https://doi.org/10.1039/d2cp01517j} {\bibfield  {journal} {\bibinfo
  {journal} {Phys. Chem. Chem. Phys.}\ }\textbf {\bibinfo {volume} {24}},\
  \bibinfo {pages} {21538--21548} (\bibinfo {year} {2022})}\BibitemShut
  {NoStop}%
\bibitem [{\citenamefont {Sim}\ \emph {et~al.}(2022)\citenamefont {Sim},
  \citenamefont {Song}, \citenamefont {Vuckovic},\ and\ \citenamefont
  {Burke}}]{Sim2022}%
  \BibitemOpen
  \bibfield  {author} {\bibinfo {author} {\bibfnamefont {E.}~\bibnamefont
  {Sim}}, \bibinfo {author} {\bibfnamefont {S.}~\bibnamefont {Song}}, \bibinfo
  {author} {\bibfnamefont {S.}~\bibnamefont {Vuckovic}},\ and\ \bibinfo
  {author} {\bibfnamefont {K.}~\bibnamefont {Burke}},\ }\bibfield  {title}
  {\enquote {\bibinfo {title} {Improving results by improving densities:
  Density-corrected density functional theory},}\ }\href
  {https://doi.org/10.1021/jacs.1c11506} {\bibfield  {journal} {\bibinfo
  {journal} {J. Am. Chem. Soc.}\ }\textbf {\bibinfo {volume} {144}},\ \bibinfo
  {pages} {6625--6639} (\bibinfo {year} {2022})}\BibitemShut {NoStop}%
\bibitem [{\citenamefont {Dasgupta}\ \emph {et~al.}(2022)\citenamefont
  {Dasgupta}, \citenamefont {Shahi}, \citenamefont {Bhetwal}, \citenamefont
  {Perdew},\ and\ \citenamefont {Paesani}}]{Dasgupta2022}%
  \BibitemOpen
  \bibfield  {author} {\bibinfo {author} {\bibfnamefont {S.}~\bibnamefont
  {Dasgupta}}, \bibinfo {author} {\bibfnamefont {C.}~\bibnamefont {Shahi}},
  \bibinfo {author} {\bibfnamefont {P.}~\bibnamefont {Bhetwal}}, \bibinfo
  {author} {\bibfnamefont {J.~P.}\ \bibnamefont {Perdew}},\ and\ \bibinfo
  {author} {\bibfnamefont {F.}~\bibnamefont {Paesani}},\ }\bibfield  {title}
  {\enquote {\bibinfo {title} {How good is the density-corrected {SCAN}
  functional for neutral and ionic aqueous systems, and what is so right about
  the hartree{\textendash}fock density?}}\ }\href
  {https://doi.org/10.1021/acs.jctc.2c00313} {\bibfield  {journal} {\bibinfo
  {journal} {J. Chem. Theory Comput.}\ }\textbf {\bibinfo {volume} {18}},\
  \bibinfo {pages} {4745--4761} (\bibinfo {year} {2022})}\BibitemShut {NoStop}%
\bibitem [{\citenamefont {Weymuth}\ and\ \citenamefont
  {Reiher}(2022)}]{Weymuth2022}%
  \BibitemOpen
  \bibfield  {author} {\bibinfo {author} {\bibfnamefont {T.}~\bibnamefont
  {Weymuth}}\ and\ \bibinfo {author} {\bibfnamefont {M.}~\bibnamefont
  {Reiher}},\ }\bibfield  {title} {\enquote {\bibinfo {title} {The
  transferability limits of static benchmarks},}\ }\href
  {https://doi.org/10.1039/d2cp01725c} {\bibfield  {journal} {\bibinfo
  {journal} {Phys. Chem. Chem. Phys.}\ }\textbf {\bibinfo {volume} {24}},\
  \bibinfo {pages} {14692--14698} (\bibinfo {year} {2022})}\BibitemShut
  {NoStop}%
\bibitem [{\citenamefont {Kim}, \citenamefont {Sim},\ and\ \citenamefont
  {Burke}(2013)}]{Kim2013-DensDriven}%
  \BibitemOpen
  \bibfield  {author} {\bibinfo {author} {\bibfnamefont {M.-C.}\ \bibnamefont
  {Kim}}, \bibinfo {author} {\bibfnamefont {E.}~\bibnamefont {Sim}},\ and\
  \bibinfo {author} {\bibfnamefont {K.}~\bibnamefont {Burke}},\ }\bibfield
  {title} {\enquote {\bibinfo {title} {Understanding and reducing errors in
  density functional calculations},}\ }\href@noop {} {\bibfield  {journal}
  {\bibinfo  {journal} {Phys. Rev. Lett.}\ }\textbf {\bibinfo {volume} {111}},\
  \bibinfo {pages} {073003} (\bibinfo {year} {2013})}\BibitemShut {NoStop}%
\bibitem [{\citenamefont {Vuckovic}\ \emph {et~al.}(2019)\citenamefont
  {Vuckovic}, \citenamefont {Song}, \citenamefont {Kozlowski}, \citenamefont
  {Sim},\ and\ \citenamefont {Burke}}]{Vuckovic2019}%
  \BibitemOpen
  \bibfield  {author} {\bibinfo {author} {\bibfnamefont {S.}~\bibnamefont
  {Vuckovic}}, \bibinfo {author} {\bibfnamefont {S.}~\bibnamefont {Song}},
  \bibinfo {author} {\bibfnamefont {J.}~\bibnamefont {Kozlowski}}, \bibinfo
  {author} {\bibfnamefont {E.}~\bibnamefont {Sim}},\ and\ \bibinfo {author}
  {\bibfnamefont {K.}~\bibnamefont {Burke}},\ }\bibfield  {title} {\enquote
  {\bibinfo {title} {Density functional analysis: The theory of
  density-corrected {DFT}},}\ }\href {https://doi.org/10.1021/acs.jctc.9b00826}
  {\bibfield  {journal} {\bibinfo  {journal} {J. Chem. Theory Comput.}\
  }\textbf {\bibinfo {volume} {15}},\ \bibinfo {pages} {6636--6646} (\bibinfo
  {year} {2019})}\BibitemShut {NoStop}%
\bibitem [{\citenamefont {Parrish}\ \emph {et~al.}(2017)\citenamefont
  {Parrish}, \citenamefont {Burns}, \citenamefont {Smith}, \citenamefont
  {Simmonett}, \citenamefont {DePrince}, \citenamefont {Hohenstein},
  \citenamefont {Bozkaya}, \citenamefont {Sokolov}, \citenamefont {Remigio},
  \citenamefont {Richard}, \citenamefont {Gonthier}, \citenamefont {James},
  \citenamefont {McAlexander}, \citenamefont {Kumar}, \citenamefont {Saitow},
  \citenamefont {Wang}, \citenamefont {Pritchard}, \citenamefont {Verma},
  \citenamefont {Schaefer}, \citenamefont {Patkowski}, \citenamefont {King},
  \citenamefont {Valeev}, \citenamefont {Evangelista}, \citenamefont {Turney},
  \citenamefont {Crawford},\ and\ \citenamefont {Sherrill}}]{psi4}%
  \BibitemOpen
  \bibfield  {author} {\bibinfo {author} {\bibfnamefont {R.~M.}\ \bibnamefont
  {Parrish}}, \bibinfo {author} {\bibfnamefont {L.~A.}\ \bibnamefont {Burns}},
  \bibinfo {author} {\bibfnamefont {D.~G.~A.}\ \bibnamefont {Smith}}, \bibinfo
  {author} {\bibfnamefont {A.~C.}\ \bibnamefont {Simmonett}}, \bibinfo {author}
  {\bibfnamefont {A.~E.}\ \bibnamefont {DePrince}}, \bibinfo {author}
  {\bibfnamefont {E.~G.}\ \bibnamefont {Hohenstein}}, \bibinfo {author}
  {\bibfnamefont {U.}~\bibnamefont {Bozkaya}}, \bibinfo {author} {\bibfnamefont
  {A.~Y.}\ \bibnamefont {Sokolov}}, \bibinfo {author} {\bibfnamefont {R.~D.}\
  \bibnamefont {Remigio}}, \bibinfo {author} {\bibfnamefont {R.~M.}\
  \bibnamefont {Richard}}, \bibinfo {author} {\bibfnamefont {J.~F.}\
  \bibnamefont {Gonthier}}, \bibinfo {author} {\bibfnamefont {A.~M.}\
  \bibnamefont {James}}, \bibinfo {author} {\bibfnamefont {H.~R.}\ \bibnamefont
  {McAlexander}}, \bibinfo {author} {\bibfnamefont {A.}~\bibnamefont {Kumar}},
  \bibinfo {author} {\bibfnamefont {M.}~\bibnamefont {Saitow}}, \bibinfo
  {author} {\bibfnamefont {X.}~\bibnamefont {Wang}}, \bibinfo {author}
  {\bibfnamefont {B.~P.}\ \bibnamefont {Pritchard}}, \bibinfo {author}
  {\bibfnamefont {P.}~\bibnamefont {Verma}}, \bibinfo {author} {\bibfnamefont
  {H.~F.}\ \bibnamefont {Schaefer}}, \bibinfo {author} {\bibfnamefont
  {K.}~\bibnamefont {Patkowski}}, \bibinfo {author} {\bibfnamefont {R.~A.}\
  \bibnamefont {King}}, \bibinfo {author} {\bibfnamefont {E.~F.}\ \bibnamefont
  {Valeev}}, \bibinfo {author} {\bibfnamefont {F.~A.}\ \bibnamefont
  {Evangelista}}, \bibinfo {author} {\bibfnamefont {J.~M.}\ \bibnamefont
  {Turney}}, \bibinfo {author} {\bibfnamefont {T.~D.}\ \bibnamefont
  {Crawford}},\ and\ \bibinfo {author} {\bibfnamefont {C.~D.}\ \bibnamefont
  {Sherrill}},\ }\bibfield  {title} {\enquote {\bibinfo {title} {Psi4 1.1: An
  open-source electronic structure program emphasizing automation, advanced
  libraries, and interoperability},}\ }\href@noop {} {\bibfield  {journal}
  {\bibinfo  {journal} {J. Chem. Theory Comput.}\ }\textbf {\bibinfo {volume}
  {13}},\ \bibinfo {pages} {3185--3197} (\bibinfo {year} {2017})}\BibitemShut
  {NoStop}%
\bibitem [{\citenamefont {Smith}\ \emph {et~al.}(2018)\citenamefont {Smith},
  \citenamefont {Burns}, \citenamefont {Sirianni}, \citenamefont {Nascimento},
  \citenamefont {Kumar}, \citenamefont {James}, \citenamefont {Schriber},
  \citenamefont {Zhang}, \citenamefont {Zhang}, \citenamefont {Abbott},
  \citenamefont {Berquist}, \citenamefont {Lechner}, \citenamefont {Cunha},
  \citenamefont {Heide}, \citenamefont {Waldrop}, \citenamefont {Takeshita},
  \citenamefont {Alenaizan}, \citenamefont {Neuhauser}, \citenamefont {King},
  \citenamefont {Simmonett}, \citenamefont {Turney}, \citenamefont {Schaefer},
  \citenamefont {Evangelista}, \citenamefont {DePrince}, \citenamefont
  {Crawford}, \citenamefont {Patkowski},\ and\ \citenamefont
  {Sherrill}}]{psi4numpy}%
  \BibitemOpen
  \bibfield  {author} {\bibinfo {author} {\bibfnamefont {D.~G.~A.}\
  \bibnamefont {Smith}}, \bibinfo {author} {\bibfnamefont {L.~A.}\ \bibnamefont
  {Burns}}, \bibinfo {author} {\bibfnamefont {D.~A.}\ \bibnamefont {Sirianni}},
  \bibinfo {author} {\bibfnamefont {D.~R.}\ \bibnamefont {Nascimento}},
  \bibinfo {author} {\bibfnamefont {A.}~\bibnamefont {Kumar}}, \bibinfo
  {author} {\bibfnamefont {A.~M.}\ \bibnamefont {James}}, \bibinfo {author}
  {\bibfnamefont {J.~B.}\ \bibnamefont {Schriber}}, \bibinfo {author}
  {\bibfnamefont {T.}~\bibnamefont {Zhang}}, \bibinfo {author} {\bibfnamefont
  {B.}~\bibnamefont {Zhang}}, \bibinfo {author} {\bibfnamefont {A.~S.}\
  \bibnamefont {Abbott}}, \bibinfo {author} {\bibfnamefont {E.~J.}\
  \bibnamefont {Berquist}}, \bibinfo {author} {\bibfnamefont {M.~H.}\
  \bibnamefont {Lechner}}, \bibinfo {author} {\bibfnamefont {L.~A.}\
  \bibnamefont {Cunha}}, \bibinfo {author} {\bibfnamefont {A.~G.}\ \bibnamefont
  {Heide}}, \bibinfo {author} {\bibfnamefont {J.~M.}\ \bibnamefont {Waldrop}},
  \bibinfo {author} {\bibfnamefont {T.~Y.}\ \bibnamefont {Takeshita}}, \bibinfo
  {author} {\bibfnamefont {A.}~\bibnamefont {Alenaizan}}, \bibinfo {author}
  {\bibfnamefont {D.}~\bibnamefont {Neuhauser}}, \bibinfo {author}
  {\bibfnamefont {R.~A.}\ \bibnamefont {King}}, \bibinfo {author}
  {\bibfnamefont {A.~C.}\ \bibnamefont {Simmonett}}, \bibinfo {author}
  {\bibfnamefont {J.~M.}\ \bibnamefont {Turney}}, \bibinfo {author}
  {\bibfnamefont {H.~F.}\ \bibnamefont {Schaefer}}, \bibinfo {author}
  {\bibfnamefont {F.~A.}\ \bibnamefont {Evangelista}}, \bibinfo {author}
  {\bibfnamefont {A.~E.}\ \bibnamefont {DePrince}}, \bibinfo {author}
  {\bibfnamefont {T.~D.}\ \bibnamefont {Crawford}}, \bibinfo {author}
  {\bibfnamefont {K.}~\bibnamefont {Patkowski}},\ and\ \bibinfo {author}
  {\bibfnamefont {C.~D.}\ \bibnamefont {Sherrill}},\ }\bibfield  {title}
  {\enquote {\bibinfo {title} {Psi4numpy: An interactive quantum chemistry
  programming environment for reference implementations and rapid
  development},}\ }\href@noop {} {\bibfield  {journal} {\bibinfo  {journal} {J.
  Chem. Theory Comput.}\ }\textbf {\bibinfo {volume} {14}},\ \bibinfo {pages}
  {3504--3511} (\bibinfo {year} {2018})}\BibitemShut {NoStop}%
\bibitem [{\citenamefont {Weigend}\ and\ \citenamefont
  {Ahlrichs}(2005)}]{Weigend2005}%
  \BibitemOpen
  \bibfield  {author} {\bibinfo {author} {\bibfnamefont {F.}~\bibnamefont
  {Weigend}}\ and\ \bibinfo {author} {\bibfnamefont {R.}~\bibnamefont
  {Ahlrichs}},\ }\bibfield  {title} {\enquote {\bibinfo {title} {Balanced basis
  sets of split valence, triple zeta valence and quadruple zeta valence quality
  for {H} to {Rn}: Design and assessment of accuracy},}\ }\href
  {https://doi.org/10.1039/b508541a} {\bibfield  {journal} {\bibinfo  {journal}
  {Phys. Chem. Chem. Phys.}\ }\textbf {\bibinfo {volume} {7}},\ \bibinfo
  {pages} {3297} (\bibinfo {year} {2005})}\BibitemShut {NoStop}%
\bibitem [{\citenamefont {Gould}(2023)}]{Gould2023}%
  \BibitemOpen
  \bibfield  {author} {\bibinfo {author} {\bibfnamefont {T.}~\bibnamefont
  {Gould}},\ }\bibfield  {title} {\enquote {\bibinfo {title} {Toward routine
  kohn{\textendash}sham inversion using the
  {\textquotedblleft}lieb-response{\textquotedblright} approach},}\ }\href
  {https://doi.org/10.1063/5.0134330} {\bibfield  {journal} {\bibinfo
  {journal} {J. Chem. Phys.}\ }\textbf {\bibinfo {volume} {158}} (\bibinfo
  {year} {2023}),\ 10.1063/5.0134330}\BibitemShut {NoStop}%
\bibitem [{\citenamefont {Vosko}\ \emph {et~al.}(1980)\citenamefont {Vosko},
  \citenamefont {Wilk}, ,\ and\ \citenamefont {Nusair}}]{g09_svwn}%
  \BibitemOpen
  \bibfield  {author} {\bibinfo {author} {\bibfnamefont {S.~H.}\ \bibnamefont
  {Vosko}}, \bibinfo {author} {\bibfnamefont {L.}~\bibnamefont {Wilk}}, ,\ and\
  \bibinfo {author} {\bibfnamefont {M.}~\bibnamefont {Nusair}},\ }\bibfield
  {title} {\enquote {\bibinfo {title} {Accurate spin-dependent electron liquid
  correlation energies for local spin density calculations: A critical
  analysis},}\ }\href@noop {} {\bibfield  {journal} {\bibinfo  {journal} {Can.
  J. Phys.}\ }\textbf {\bibinfo {volume} {58}},\ \bibinfo {pages} {1200--11}
  (\bibinfo {year} {1980})}\BibitemShut {NoStop}%
\bibitem [{\citenamefont {Perdew}\ \emph {et~al.}(1996)\citenamefont {Perdew},
  \citenamefont {Burke}, ,\ and\ \citenamefont {Ernzerhof}}]{g09_pbepbe}%
  \BibitemOpen
  \bibfield  {author} {\bibinfo {author} {\bibfnamefont {J.~P.}\ \bibnamefont
  {Perdew}}, \bibinfo {author} {\bibfnamefont {K.}~\bibnamefont {Burke}}, ,\
  and\ \bibinfo {author} {\bibfnamefont {M.}~\bibnamefont {Ernzerhof}},\
  }\bibfield  {title} {\enquote {\bibinfo {title} {Generalized gradient
  approximation made simple},}\ }\href@noop {} {\bibfield  {journal} {\bibinfo
  {journal} {Phys. Rev. Lett.}\ }\textbf {\bibinfo {volume} {77}},\ \bibinfo
  {pages} {3865--68} (\bibinfo {year} {1996})}\BibitemShut {NoStop}%
\bibitem [{\citenamefont {Becke}(1988)}]{g09_blyp}%
  \BibitemOpen
  \bibfield  {author} {\bibinfo {author} {\bibfnamefont {A.~D.}\ \bibnamefont
  {Becke}},\ }\bibfield  {title} {\enquote {\bibinfo {title}
  {Density-functional exchange-energy approximation with correct
  asymptotic-behavior},}\ }\href@noop {} {\bibfield  {journal} {\bibinfo
  {journal} {Phys. Rev. A}\ }\textbf {\bibinfo {volume} {38}},\ \bibinfo
  {pages} {3098--100} (\bibinfo {year} {1988})}\BibitemShut {NoStop}%
\bibitem [{\citenamefont {Peverati}\ and\ \citenamefont
  {Truhlar}(2012)}]{g09_N12}%
  \BibitemOpen
  \bibfield  {author} {\bibinfo {author} {\bibfnamefont {R.}~\bibnamefont
  {Peverati}}\ and\ \bibinfo {author} {\bibfnamefont {D.~G.}\ \bibnamefont
  {Truhlar}},\ }\bibfield  {title} {\enquote {\bibinfo {title}
  {Exchange-correlation functional with good accuracy for both structural and
  energetic properties while depending only on the density and its gradient},}\
  }\href@noop {} {\bibfield  {journal} {\bibinfo  {journal} {J. Chem. Theory
  and Comput.}\ }\textbf {\bibinfo {volume} {8}},\ \bibinfo {pages}
  {2310--2319} (\bibinfo {year} {2012})}\BibitemShut {NoStop}%
\bibitem [{\citenamefont {Peverati}, \citenamefont {Zhao},\ and\ \citenamefont
  {Truhlar}(2011)}]{g09_SOGGA11}%
  \BibitemOpen
  \bibfield  {author} {\bibinfo {author} {\bibfnamefont {R.}~\bibnamefont
  {Peverati}}, \bibinfo {author} {\bibfnamefont {Y.}~\bibnamefont {Zhao}},\
  and\ \bibinfo {author} {\bibfnamefont {D.~G.}\ \bibnamefont {Truhlar}},\
  }\bibfield  {title} {\enquote {\bibinfo {title} {Generalized gradient
  approximation that recovers the second-order density-gradient expansion with
  optimized across-the-board performance},}\ }\href@noop {} {\bibfield
  {journal} {\bibinfo  {journal} {J. Phys. Chem. Lett.}\ }\textbf {\bibinfo
  {volume} {2}},\ \bibinfo {pages} {1991--1997} (\bibinfo {year}
  {2011})}\BibitemShut {NoStop}%
\bibitem [{\citenamefont {Tao}\ \emph {et~al.}(2003)\citenamefont {Tao},
  \citenamefont {Perdew}, \citenamefont {Staroverov}, ,\ and\ \citenamefont
  {Scuseria}}]{g09_tpsstpss}%
  \BibitemOpen
  \bibfield  {author} {\bibinfo {author} {\bibfnamefont {J.~M.}\ \bibnamefont
  {Tao}}, \bibinfo {author} {\bibfnamefont {J.~P.}\ \bibnamefont {Perdew}},
  \bibinfo {author} {\bibfnamefont {V.~N.}\ \bibnamefont {Staroverov}}, ,\ and\
  \bibinfo {author} {\bibfnamefont {G.~E.}\ \bibnamefont {Scuseria}},\
  }\bibfield  {title} {\enquote {\bibinfo {title} {Climbing the density
  functional ladder: Nonempirical meta-generalized gradient approximation
  designed for molecules and solids},}\ }\href@noop {} {\bibfield  {journal}
  {\bibinfo  {journal} {Phys. Rev. Lett.}\ }\textbf {\bibinfo {volume} {91}},\
  \bibinfo {pages} {146401} (\bibinfo {year} {2003})}\BibitemShut {NoStop}%
\bibitem [{\citenamefont {Sun}, \citenamefont {Ruzsinszky},\ and\ \citenamefont
  {Perdew}(2015)}]{DFA:SCAN}%
  \BibitemOpen
  \bibfield  {author} {\bibinfo {author} {\bibfnamefont {J.}~\bibnamefont
  {Sun}}, \bibinfo {author} {\bibfnamefont {A.}~\bibnamefont {Ruzsinszky}},\
  and\ \bibinfo {author} {\bibfnamefont {J.}~\bibnamefont {Perdew}},\
  }\bibfield  {title} {\enquote {\bibinfo {title} {Strongly constrained and
  appropriately normed semilocal density functional},}\ }\href
  {https://doi.org/10.1103/physrevlett.115.036402} {\bibfield  {journal}
  {\bibinfo  {journal} {Physical Review Letters}\ }\textbf {\bibinfo {volume}
  {115}},\ \bibinfo {pages} {036402} (\bibinfo {year} {2015})}\BibitemShut
  {NoStop}%
\bibitem [{\citenamefont {Zhao}\ and\ \citenamefont
  {Truhlar}(2006)}]{g09_M06L}%
  \BibitemOpen
  \bibfield  {author} {\bibinfo {author} {\bibfnamefont {Y.}~\bibnamefont
  {Zhao}}\ and\ \bibinfo {author} {\bibfnamefont {D.~G.}\ \bibnamefont
  {Truhlar}},\ }\bibfield  {title} {\enquote {\bibinfo {title} {A new local
  density functional for main-group thermochemistry, transition metal bonding,
  thermochemical kinetics, and noncovalent interactions},}\ }\href@noop {}
  {\bibfield  {journal} {\bibinfo  {journal} {J. Chem. Phys.}\ }\textbf
  {\bibinfo {volume} {125}},\ \bibinfo {pages} {194101} (\bibinfo {year}
  {2006})}\BibitemShut {NoStop}%
\bibitem [{\citenamefont {Yu}\ \emph {et~al.}(2016)\citenamefont {Yu},
  \citenamefont {He}, ,\ and\ \citenamefont {Truhlar}}]{g09_MN15L}%
  \BibitemOpen
  \bibfield  {author} {\bibinfo {author} {\bibfnamefont {H.~S.}\ \bibnamefont
  {Yu}}, \bibinfo {author} {\bibfnamefont {X.}~\bibnamefont {He}}, ,\ and\
  \bibinfo {author} {\bibfnamefont {D.~G.}\ \bibnamefont {Truhlar}},\
  }\bibfield  {title} {\enquote {\bibinfo {title} {Mn15-l: A new local
  exchange-correlation functional for kohn–sham density functional theory
  with broad accuracy for atoms, molecules, and solids},}\ }\href@noop {}
  {\bibfield  {journal} {\bibinfo  {journal} {J. Chem. Theory and Comput.}\
  }\textbf {\bibinfo {volume} {12}},\ \bibinfo {pages} {1280--1293} (\bibinfo
  {year} {2016})}\BibitemShut {NoStop}%
\bibitem [{\citenamefont {Adamo}\ and\ \citenamefont
  {Barone}(1999)}]{g09_pbe1pbe}%
  \BibitemOpen
  \bibfield  {author} {\bibinfo {author} {\bibfnamefont {C.}~\bibnamefont
  {Adamo}}\ and\ \bibinfo {author} {\bibfnamefont {V.}~\bibnamefont {Barone}},\
  }\bibfield  {title} {\enquote {\bibinfo {title} {Toward reliable density
  functional methods without adjustable parameters: The pbe0 model},}\
  }\href@noop {} {\bibfield  {journal} {\bibinfo  {journal} {J. Chem. Phys.}\
  }\textbf {\bibinfo {volume} {110}},\ \bibinfo {pages} {6158--69} (\bibinfo
  {year} {1999})}\BibitemShut {NoStop}%
\bibitem [{\citenamefont {Becke}(1993)}]{g09_b3lyp}%
  \BibitemOpen
  \bibfield  {author} {\bibinfo {author} {\bibfnamefont {A.~D.}\ \bibnamefont
  {Becke}},\ }\bibfield  {title} {\enquote {\bibinfo {title} {A new mixing of
  hartree-fock and local density-functional theories},}\ }\href@noop {}
  {\bibfield  {journal} {\bibinfo  {journal} {J. Chem. Phys.}\ }\textbf
  {\bibinfo {volume} {98}},\ \bibinfo {pages} {1372--77} (\bibinfo {year}
  {1993})}\BibitemShut {NoStop}%
\bibitem [{\citenamefont {Zhao}\ and\ \citenamefont {Truhlar}(2008)}]{g09_m06}%
  \BibitemOpen
  \bibfield  {author} {\bibinfo {author} {\bibfnamefont {Y.}~\bibnamefont
  {Zhao}}\ and\ \bibinfo {author} {\bibfnamefont {D.~G.}\ \bibnamefont
  {Truhlar}},\ }\bibfield  {title} {\enquote {\bibinfo {title} {The m06 suite
  of density functionals for main group thermochemistry, thermochemical
  kinetics, noncovalent interactions, excited states, and transition elements:
  two new functionals and systematic testing of four m06-class functionals and
  12 other functionals},}\ }\href@noop {} {\bibfield  {journal} {\bibinfo
  {journal} {Theor. Chem. Acc.}\ }\textbf {\bibinfo {volume} {120}},\ \bibinfo
  {pages} {215--41} (\bibinfo {year} {2008})}\BibitemShut {NoStop}%
\bibitem [{\citenamefont {Chai}\ and\ \citenamefont
  {Head-Gordon}(2008)}]{g09_wB97x}%
  \BibitemOpen
  \bibfield  {author} {\bibinfo {author} {\bibfnamefont {J.-D.}\ \bibnamefont
  {Chai}}\ and\ \bibinfo {author} {\bibfnamefont {M.}~\bibnamefont
  {Head-Gordon}},\ }\bibfield  {title} {\enquote {\bibinfo {title} {Systematic
  optimization of long-range corrected hybrid density functionals},}\
  }\href@noop {} {\bibfield  {journal} {\bibinfo  {journal} {J. Chem. Phys.}\
  }\textbf {\bibinfo {volume} {128}},\ \bibinfo {pages} {084106} (\bibinfo
  {year} {2008})}\BibitemShut {NoStop}%
\bibitem [{\citenamefont {Grimme}(2006)}]{g09_b2plyp}%
  \BibitemOpen
  \bibfield  {author} {\bibinfo {author} {\bibfnamefont {S.}~\bibnamefont
  {Grimme}},\ }\bibfield  {title} {\enquote {\bibinfo {title} {Semiempirical
  hybrid density functional with perturbative second-order correlation},}\
  }\href@noop {} {\bibfield  {journal} {\bibinfo  {journal} {J. Chem. Phys.}\
  }\textbf {\bibinfo {volume} {124}},\ \bibinfo {pages} {03410} (\bibinfo
  {year} {2006})}\BibitemShut {NoStop}%
\bibitem [{Note1()}]{Note1}%
  \BibitemOpen
  \bibinfo {note} {Twenty of each were initially computed, but cases where one
  or more DFA did not converge where excluded from the final list.}\BibitemShut
  {Stop}%
\bibitem [{\citenamefont {Song}\ \emph {et~al.}(2021)\citenamefont {Song},
  \citenamefont {Vuckovic}, \citenamefont {Sim},\ and\ \citenamefont
  {Burke}}]{Song2021}%
  \BibitemOpen
  \bibfield  {author} {\bibinfo {author} {\bibfnamefont {S.}~\bibnamefont
  {Song}}, \bibinfo {author} {\bibfnamefont {S.}~\bibnamefont {Vuckovic}},
  \bibinfo {author} {\bibfnamefont {E.}~\bibnamefont {Sim}},\ and\ \bibinfo
  {author} {\bibfnamefont {K.}~\bibnamefont {Burke}},\ }\bibfield  {title}
  {\enquote {\bibinfo {title} {Density sensitivity of empirical functionals},}\
  }\href {https://doi.org/10.1021/acs.jpclett.0c03545} {\bibfield  {journal}
  {\bibinfo  {journal} {J. Phys. Chem. Lett.}\ }\textbf {\bibinfo {volume}
  {12}},\ \bibinfo {pages} {800--807} (\bibinfo {year} {2021})}\BibitemShut
  {NoStop}%
\bibitem [{\citenamefont {Wang}\ and\ \citenamefont {Parr}(1993)}]{Wang1993}%
  \BibitemOpen
  \bibfield  {author} {\bibinfo {author} {\bibfnamefont {Y.}~\bibnamefont
  {Wang}}\ and\ \bibinfo {author} {\bibfnamefont {R.~G.}\ \bibnamefont
  {Parr}},\ }\bibfield  {title} {\enquote {\bibinfo {title} {Construction of
  exact {Kohn-Sham} orbitals from a given electron density},}\ }\href@noop {}
  {\bibfield  {journal} {\bibinfo  {journal} {Phys. Rev. A}\ }\textbf {\bibinfo
  {volume} {47}},\ \bibinfo {pages} {R1591} (\bibinfo {year}
  {1993})}\BibitemShut {NoStop}%
\bibitem [{\citenamefont {Wu}\ and\ \citenamefont {Yang}(2003)}]{Wu2003}%
  \BibitemOpen
  \bibfield  {author} {\bibinfo {author} {\bibfnamefont {Q.}~\bibnamefont
  {Wu}}\ and\ \bibinfo {author} {\bibfnamefont {W.}~\bibnamefont {Yang}},\
  }\bibfield  {title} {\enquote {\bibinfo {title} {A direct optimization method
  for calculating density functionals and exchange{\textendash}correlation
  potentials from electron densities},}\ }\href
  {https://doi.org/10.1063/1.1535422} {\bibfield  {journal} {\bibinfo
  {journal} {J. Chem. Phys.}\ }\textbf {\bibinfo {volume} {118}},\ \bibinfo
  {pages} {2498--2509} (\bibinfo {year} {2003})}\BibitemShut {NoStop}%
\bibitem [{\citenamefont {Kanungo}, \citenamefont {Zimmerman},\ and\
  \citenamefont {Gavini}(2019)}]{Kanungo2019}%
  \BibitemOpen
  \bibfield  {author} {\bibinfo {author} {\bibfnamefont {B.}~\bibnamefont
  {Kanungo}}, \bibinfo {author} {\bibfnamefont {P.~M.}\ \bibnamefont
  {Zimmerman}},\ and\ \bibinfo {author} {\bibfnamefont {V.}~\bibnamefont
  {Gavini}},\ }\bibfield  {title} {\enquote {\bibinfo {title} {Exact
  exchange-correlation potentials from ground-state electron densities},}\
  }\href {https://doi.org/10.1038/s41467-019-12467-0} {\bibfield  {journal}
  {\bibinfo  {journal} {Nat. Commun.}\ }\textbf {\bibinfo {volume} {10}}
  (\bibinfo {year} {2019}),\ 10.1038/s41467-019-12467-0}\BibitemShut {NoStop}%
\bibitem [{\citenamefont {Shi}\ and\ \citenamefont
  {Wasserman}(2021)}]{Shi2021}%
  \BibitemOpen
  \bibfield  {author} {\bibinfo {author} {\bibfnamefont {Y.}~\bibnamefont
  {Shi}}\ and\ \bibinfo {author} {\bibfnamefont {A.}~\bibnamefont
  {Wasserman}},\ }\bibfield  {title} {\enquote {\bibinfo {title} {Inverse
  kohn-sham density functional theory: Progress and challenges},}\ }\href
  {https://doi.org/10.1021/acs.jpclett.1c00752} {\bibfield  {journal} {\bibinfo
   {journal} {J. Phys. Chem. Lett.}\ }\textbf {\bibinfo {volume} {12}},\
  \bibinfo {pages} {5308--5318} (\bibinfo {year} {2021})}\BibitemShut {NoStop}%
\bibitem [{\citenamefont {Kanungo}\ \emph {et~al.}(2023)\citenamefont
  {Kanungo}, \citenamefont {Kaplan}, \citenamefont {Shahi}, \citenamefont
  {Gavini},\ and\ \citenamefont {Perdew}}]{Kanungo2023}%
  \BibitemOpen
  \bibfield  {author} {\bibinfo {author} {\bibfnamefont {B.}~\bibnamefont
  {Kanungo}}, \bibinfo {author} {\bibfnamefont {A.~D.}\ \bibnamefont {Kaplan}},
  \bibinfo {author} {\bibfnamefont {C.}~\bibnamefont {Shahi}}, \bibinfo
  {author} {\bibfnamefont {V.}~\bibnamefont {Gavini}},\ and\ \bibinfo {author}
  {\bibfnamefont {J.~P.}\ \bibnamefont {Perdew}},\ }\href
  {https://doi.org/10.48550/ARXIV.2303.05318} {\enquote {\bibinfo {title}
  {Unconventional error cancellation explains the success of hartree-fock
  density functional theory for barrier heights},}\ } (\bibinfo {year}
  {2023})\BibitemShut {NoStop}%
\bibitem [{\citenamefont {Karton}\ \emph {et~al.}(2006)\citenamefont {Karton},
  \citenamefont {Rabinovich}, \citenamefont {Martin},\ and\ \citenamefont
  {Ruscic}}]{Karton2006}%
  \BibitemOpen
  \bibfield  {author} {\bibinfo {author} {\bibfnamefont {A.}~\bibnamefont
  {Karton}}, \bibinfo {author} {\bibfnamefont {E.}~\bibnamefont {Rabinovich}},
  \bibinfo {author} {\bibfnamefont {J.~M.~L.}\ \bibnamefont {Martin}},\ and\
  \bibinfo {author} {\bibfnamefont {B.}~\bibnamefont {Ruscic}},\ }\bibfield
  {title} {\enquote {\bibinfo {title} {W4 theory for computational
  thermochemistry: In pursuit of confident sub-{kJ}/mol predictions},}\ }\href
  {https://doi.org/10.1063/1.2348881} {\bibfield  {journal} {\bibinfo
  {journal} {J. Chem. Phys.}\ }\textbf {\bibinfo {volume} {125}} (\bibinfo
  {year} {2006}),\ 10.1063/1.2348881}\BibitemShut {NoStop}%
\end{thebibliography}%

\end{document}